\documentclass[12pt,a4paper]{article}
\usepackage[utf8]{inputenc}
\usepackage{amsmath}
\usepackage{amsfonts}
\usepackage{amssymb}
\usepackage{graphics}
\usepackage{epsfig}
\usepackage[section]{placeins}

\usepackage{geometry}
\begin{document}

\title{A fundamental frequency estimation method for tonal sounds inspired on bird song studies}%
\author{C. Jarne$^*$ }
\maketitle
\begin{center}


$*$ Department of Science and Technology from University of Quilmes (UNQ) and IFIBA-CONICET, 1428 Buenos Aires, Argentina. \\e-mail: cecilia.jarne@unq.edu.ar\\
\end{center}

\begin{abstract}

A fast implementation of fundamental frequency estimation is presented in this work. The algorithm is based on a frequency-domain approach. It was mainly develop for tonal sounds and used in Canary bird song analysis. The method was implemented but not restricted for this kind of data. It could be easily adapted for other proposes. Python libraries were used to develop a code with a simple algorithm to obtain fundamental frequency. A simple open source code is provided in the local university repository.

\end{abstract}


\section{Introduction}

\subsection{Regarding the importance of fundamental frequency of animal sounds}

In the field of biology the study of animal sounds is a key to understand behavior, evolution and the differences across the animal species. In a diversity of studies the knowledge of fundamental frequency allows to infer information of the animal communication characteristics of the different clades. 

Of particular interest is the case of bird songs. In past years birdsong has turned into a very interesting problem for the scientific community. The reason is that, there are approximately 10000 species of birds known to exist, where 4000 share with humans (and just a few other examples of animals) that the acquisition of vocalization requires a certain degree of exposure to a tutor. Hundreds of studies have focused on localizing the regions in the brain involved in the learning and production of the song. The hope is to understand through this example the mechanisms involved in the acquisition of a general complex behavior through learning \cite{libro-gabo}. In bird songs fundamental frequency (called $f_0$) is one key on the study of peripheral mechanisms of vocalization. 

Beside the learning process of songbirds, there are other interesting examples where fundamental frequency is used. For instance to analyze behavioral data. One example consist in the study of a daily oscillation in the fundamental frequency of measured on Zebra Finch song that could reveal new insights into how time of day (circadian rithms) affects the ability to accomplish a variety of complex learned motor skills \cite{cita-plos}. 

In other species, such as sea mammals a remarkable example is the use of fundamental frequency in the study of acoustic signals and the supposed spoken language of the dolphins \cite{aplicacion-04}.

In the human case one goal in many speech analysis applications is to follow fast variations in the fundamental frequency ($F_0$) of a signal. Again, several studies were conducted in this field  \cite{method-01,method-04,method-06,method-07}. 

All cases presented have in common the need to calculate accurately the fundamental frequency as part a main par of the studies. The basic problem consist of extract the fundamental frequency $f_0$ from a sound signal, which is usually the lowest frequency component, or partial, that relates well to most of the other partials. In the case of a  periodic waveform, most partials are harmonically related. The frequency of this lowest partial is $f_0$ of the waveform \cite{report}.

\subsection{On the current techniques}

Several implementations of fundamental frequency were applied in a variety of studies going from music analysis to power line stability and can be found in literature \cite{aplicacion-02,aplicacion-01,aplicacion-03}.

To perform this task there are currently a lot of methods. In the case of speech, many pitch detection algorithms (PDAs) analyze a speech signal by partitioning it into segments and calculating the respective fundamental frequencies (short-term analysis) \cite{method-01}.

On the other hand, several methods have been proposed to obtain reliable $f_0$-trajectories from harmonic signals. In this sense it is possible to classify fundamental frequency trackers into five general categories: autocorrelation, adaptive filter, time domain, models of the human ears and frequency domain \cite{method-05}.

In general the autocorrelation algorithms consist of taking the correlation of a waveform with itself. One would expect exact similarity at a time lag of zero, with increasing dissimilarity as the time lag increases. Periodic waveforms exhibit an interesting autocorrelation characteristic: the autocorrelation function itself is periodic. As the time lag increases to half of the period of the waveform, the correlation decreases to a minimum. This is because the waveform is out of phase with its time-delayed copy. As the time lag increases again to the length of one period, the autocorrelation again increases back to a maximum, because the waveform and its time-delayed copy are in phase. The first peak in the autocorrelation indicates the period of the waveform. These kind of methods are most efficient at mid to low frequencies. 

The difficulty with autocorrelation techniques has been that peaks occur at sub-harmonics as well, and it is sometimes difficult to determine which peak is the fundamental frequency and which represent harmonics or partials. A method call YIN attempts to solve these problems by in several ways \cite{yin}. YIN is based on the difference function, which, while similar to autocorrelation, attempts to minimize the difference between the waveform and its delayed duplicate instead of maximizing the product (for autocorrelation).

Considering the adaptive filter methods, on pitch detector for example, is based on the analysis of the
difference between the filter output and the filter input. This difference must be close to zero. The bandpass
filter center frequency is controlled by this difference \cite{CITA-state-lane}.\\

Regarding the time domain methods, one type of pitch detector is based of the analysis of the zerocrossing
points. Preprocessing by filters has to be performed, in order to solve the problem of the low-amplitude zerocrossings caused by high-frequency components \cite{cita-state-moorer,cita-state-hermes}.\\


With respect to pitch detectors in the frequency domain, most of them are based on the analysis of the FFT spectrum, or of the cepstrum \cite{state-Schafler}. Cepstrum analysis is a form of spectral analysis where the output is the Fourier transform of the log of the magnitude spectrum of the input waveform [9]. The word cepstrum comes from reversing the first four letters in the word “spectrum”, indicating a modified spectrum. The theory behind this method relies on the fact that the Fourier transform of a pitched signal usually has a number of regularly spaced peaks, representing the harmonic spectrum of the signal. When the log magnitude of a spectrum is taken, these peaks are reduced, their amplitude brought into a usable scale, and the result is a periodic waveform in the frequency domain, the period of which (the distance between the peaks) is related to the fundamental frequency of the original signal. The Fourier transform of this waveform has a peak at the period of the original waveform \cite{report}.

Fundamental frequency estimation is still a difficult topic in audio signal processing. Many context-specific attempts have been made, and many of them work well in their specific context, but it has been difficult to develop a “context-free” $f_0$ estimator. Also, most of current tools are not develop in open source code. In addition $f_0$ estimators developed for a particular application, such as musical note detection or speech analysis, are well understood, but depend on the domain of the data: a detector designed for one domain is less accurate when applied to a different domain. The result is that currently there are many $f_0$ estimators, but few of them are appropriate to more than one domain \cite{report} and they are not open source tools. An example of a programing tool  in open source code is called audiobio \cite{audiobio}, but main focus of this tool is the use in music or speech.

Therefore, choosing a $f_0$ estimator for a animal sounds or song discrimination was a difficult task. This is the reason that motivated the work. The started point was the need of a simple $f_0$ tracking algorithm, easy to implement for mostly tonal songs with syllables of different time scale and a frequency range between $800$ $Hz$ and $8$ $KHz$. This range cover the canary (\textit{Serinus canaria}) singing range.

A simple and heuristic method based on frequency-domain approaches is proposed to estimate fundamental frequency, with the main characteristic of the used of open source and optimized python libraries. The algorithm and the software implementation were designed to be used for any researcher (particularly biologist) without the need of been a signal analysis expert or deep programing knowledge.

\section{The method}

The proposed algorithm is very straightforward. First a digital audio signal is selected to perform this task. The signal spectrogram function is used over the consecutive signal segments of a preferred length. A spectrogram is a visual representation of the spectrum of frequencies in a sound or other signal as they vary with time. In this representation each point has a given value for frequency, and intensity across different time bins. In general a colored scale  (or grayscale) is used to represent the intensity of each frequency value in each time column. 

\begin{figure}[hbt!]
\begin{center}
\includegraphics[width=15.5cm]{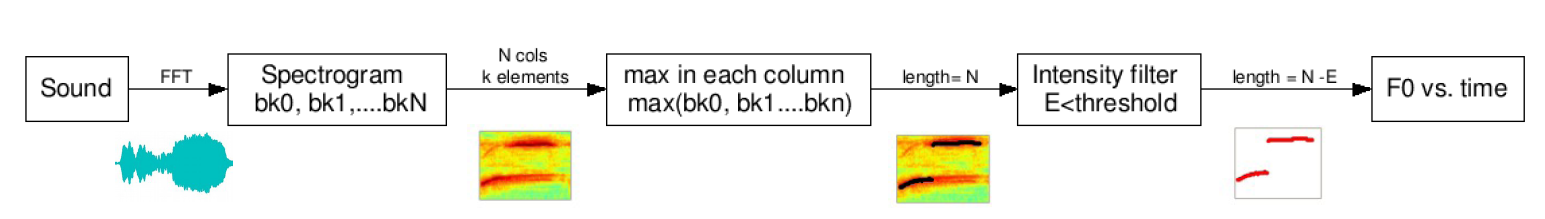}
\begin{tiny}
\caption{Algorithm Schema. Three simple steeps to obtain frequency vs. time for the original sound.}
\end{tiny}
\label{fig_00}
\end{center}
\end{figure}

The main idea is to use the appropriate resolution to estimate the sonogram with a compromise between time resolution and frequency resolution. The spectrogram is used to iteratively determinate the value with the maximum amplitude for each column across each time bin. Then only the value of maximum amplitude (more intensity in the colored scale) is saved as a new variable in each temporal bin. At this point it is possible to use the envelope audio level as a threshold to discard silence parts of the signal where no sound is produced by the bird in the data. A simple schema of the algorithm is shown in Figure \ref{fig_00}.

There is still an issue to solve. Without more intervention one have the problem of keeping the maximum of all temporal bins, but in some bins there is not signal representing a sound, there is noise or silent level. A way to get rid of the noise and keep only the values representing the desired sound is to used a second intensity filter to consider only sounds that exceed a certain threshold.  An additional consideration is needed if we want to keep fundamental frequency when harmonics are more intense that $f_0$. In this case one can use a filter over the bins to keep only frequency values between a certain range.

\begin{figure}[hbt!]
\begin{center}
\hspace*{-0.5cm}\includegraphics[totalheight=15.5cm]{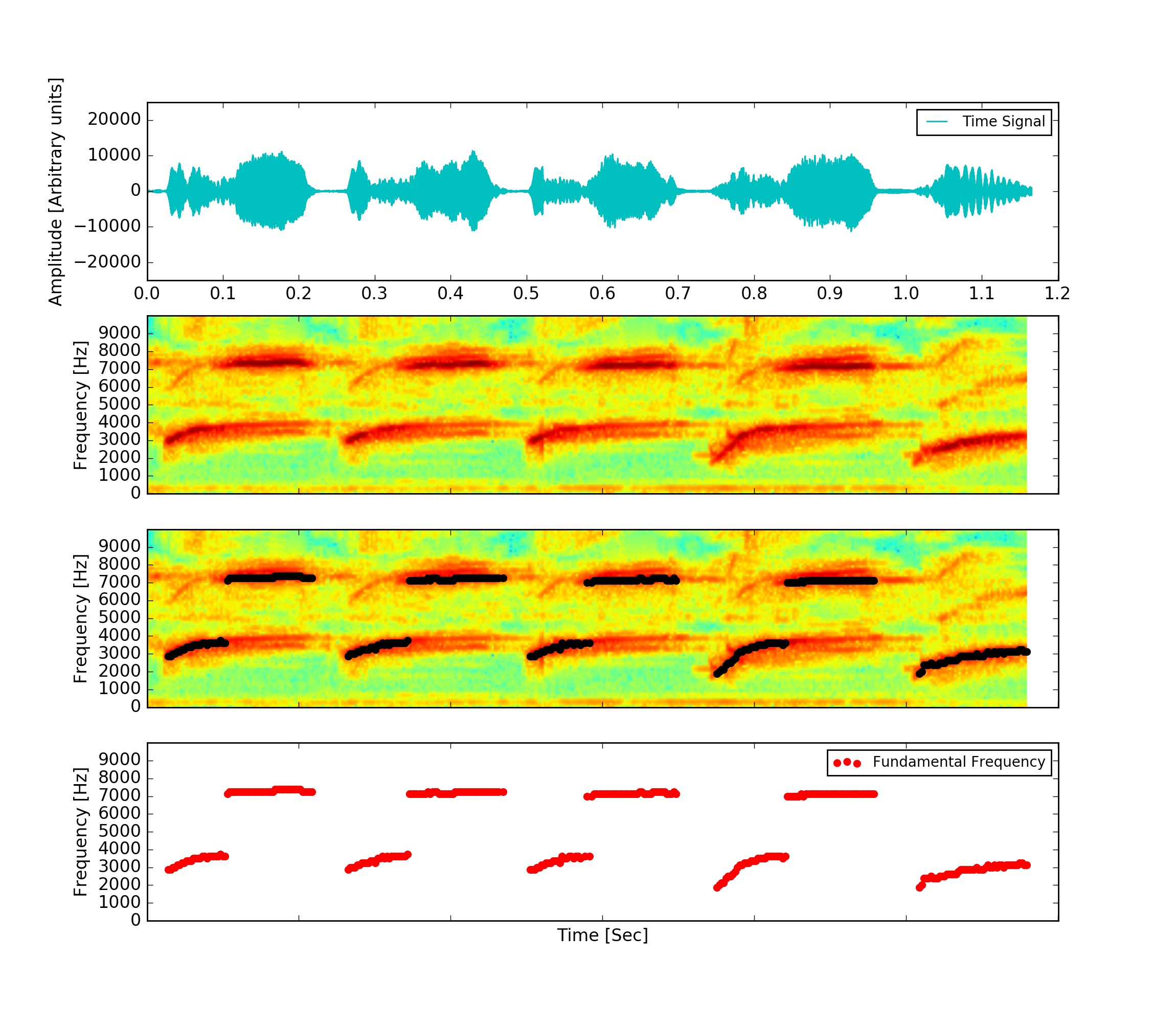}
\begin{tiny}
\hspace*{-0.75cm}
\caption{An example of the fundamental frequency estimation applied on canary sounds is shown. First panel: Sound amplitude of a singing segment of a \textit{Serinus canaria} song. Second panel: Spectrogram. Third panel: Algorithm applied over the sound segment. Fourth panel: fundamental frequency vs. time. }
\end{tiny}
\label{fig_01}
\end{center}
\end{figure}

A sound segment is used as an example to ilustrate the process. The spectrogram and it is shown in the first panel of Figure \ref{fig_01}. In this case it corresponds to a singing segment of a \textit{Serinus canaria} song. It has sample rate of $44.100$ $KHz$ and arbitrary length of $1.15$ $Sec$. The resolution of the spectrum is selected to be appropriate to the time scale of the any kind of syllables of the Canary song. Syllabic rate in canaries is from $3$ $Hz$ to $30$ $Hz$. Time resolution has to be $<\frac{1}{30}=33$ $mSec$ to distinguish between different syllables, but also smaller than the duration of the shorter syllables that is $10 mSec$.

In this way taking into account the time scale related to the sound variations, the frequency content,  the frequency range, or other characteristics, the filters over intensity threshold can be used to keep the meaningful part of the signal. 

\section{Software implementation}

The software produced in this work was developed in Python. The main reason is that Python is a free wide spread and open source programming language, where one can combine free and open-source math libraries such as Numpy or Scipy, two fundamental packages for scientific computing. The code develop is provided to be used and modified by anyone. The implementation works using the function $specgram$ in the matplotlib libraries. It takes as argument the sound vector, the windows size, sample rate and the number of points to overlap between segments; and returns the array of amplitude, sample frequencies and time bins via a Fast Fourier Transform. All these parameters can be tune according to the particular application.

An appropriate amount of overlap in the sonogram part  will depend on the choice of window and on your requirements. In contrast to welch's method, where the entire data stream is averaged over, one may wish to use a smaller overlap (or perhaps none at all) when computing a spectrogram, to maintain some statistical independence between individual segments \cite{python-03}.

After calling $specgram$ function, a filter over the frequency bins in the sonogram is applied depending of the desired frequency range. In this stage the Numpy function $argmax$ is called for each temporal bin contain the intensity values and frequencies. The value with maximum intensity is saved in a new vector. 

The implementation is a very straightforward and code is very short. It is provided in the University local repository ($http://ceciliajarne.web.unq.edu.ar/investigacion/$) with password "fundamental". It is possible to used the python 2 or modify the code for python 3. Also an output of the signal plots in the required time scale is provided  in .jpg format together with a table output of the frequency vs time values in .txt format.

\begin{figure}[hbt!]
\begin{center}
\hspace*{-0.5cm}\includegraphics[totalheight=10cm]{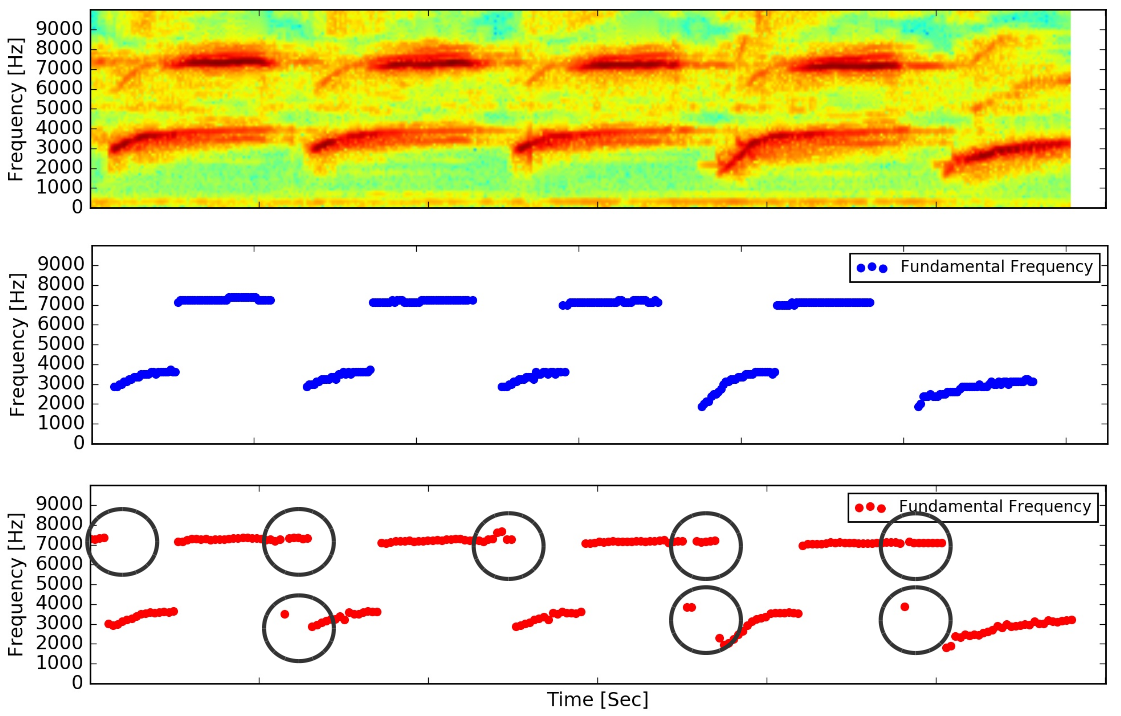}
\caption{Audio segment corresponding to \textit{Serinus canaria}. First panel: Song spectrogram. Second panel: Algorithm develop in this work applied over the sound segment. Third panel: Naive implementation to obtain fundamental frequency vs. time from reference. The bubbles show the problem of not be able to filter noise.}
\label{fig_02}
\end{center}
\end{figure}

The algorithm is very fast the implementation easy. For instance it took $12.3$ $Sec$ to obtain fundamental frequency in a signal of 20 second duration including the creation of the corresponding plots and .txt files. It only uses the standard libraries Scypi, Numpy and Matplotlib. We compared the algorithm with a simple Python implementation with a version found in \cite{stack} also written in python. That implementation lacks of proper documentation and also has no support for stereo files. It used also and additional library called Pyaudio \cite{pyaudio} to perform the task. Results obtained with both are similar but it has no clear mechanism to filter properly silence parts of the recording. They are shown in Figure \ref{fig_02}.
In this way, here is presented a simple and tunable solution to be applied for different data analysis.

\section{Conclusions} \label{conclu}

This work was motivated by the search for alternatives to obtain fundamental frequency with a simple method. To perform that task a review of current methods was perform. Here is presented in order to introduce algorithm implemented in the open source code. For tonal birdsong the algorithm is very accurate. It can be used for other animal sounds or other kind of signals, but it has the limitation of the richness of the spectrum or the noise. 

An special consideration is the case when the fundamental frequency is different than the maximum intensity value. For an initial analysis this  code could works as a interesting starting point that can be combined with other strategies to isolate and track the variations of the desired $f_0$. 

The principal advantage of this software is that a documentation of the code is provided as with clear statement of how work each part. Also data analysis is moving in the open source and collaborative software developments.
Further work could include methods to isolate fundamental frequency for more complex cases.

\section*{Funding}

This work was supported by the following institutions: CONICET and UNQ. The software was developed on python language. Code is available at University local repository:\\
$http://ceciliajarne.web.unq.edu.ar/investigacion/$ with password fundamental.

\section*{Acknowledgments}

I want to thank to the people from \textit{Laboratorio de Sistemas din\'amicos} from IFIBA institute. I want thank also the Stack overflow community also for the interesting . Finally I want to thank to Gabriel Lio for his personal help.  

\end{document}